\documentclass[a4paper]{jpconf}

\usepackage{amsmath}
\usepackage{amssymb}
\usepackage[dvips]{epsfig}
\usepackage{float}
\usepackage{graphicx}

\newcommand{\be}{\begin{equation}}
\newcommand{\ee}{\end{equation}}
\newcommand{\bea}{\begin{eqnarray}}
\newcommand{\eea}{\end{eqnarray}}

\def\ai		{{\em ab--initio}\,}

\def\pp		{{\bf p}}
\def\rr		{{\bf r}}

\def\GG		{{\bf G}}

\def\RR		{{\bf R}}

\def\qq		{{\bf q}}
\def\kk		{{\bf k}}

\def\intrp	{\int\,d{\bf r}'\,}

\newcommand{\uu}{{\bf u}}
\newcommand{\hh}{{\bf h}}

\newcommand{\SiS}{{\bf \Sigma}}
\newcommand{\bSS}{{\bf S}}
\newcommand{\VV}{{\bf V}}
\newcommand{\UU}{{\bf U}}
\newcommand{\w}{\omega}
\newcommand{\ooe}{out--of--equilibrium\,}

\def\ga         {\alpha}
\def\gb         {\beta}
\def\gc         {\gamma}

\def\gd         {\delta}

\def\gee        {\epsilon}
\def\gl         {\lambda}

\def\go         {\omega}

\def\goql       {\omega_{\qq \gl}}
\def\gql        {\qq \gl}

\def\gr         {\rho}

\def\gt         {\theta}
\def\gu         {\tau}

\def\capo       {\right.\\ \left.}

\def\rar        {\rightarrow}  
\def\la         {\langle}
\def\ra         {\rangle}
\def\dg         {\dagger}

\def\lb         {\left|}
\def\rb         {\right|}
\renewcommand{\[}{\left[}
\renewcommand{\]}{\right]}
\renewcommand{\(}{\left(}
\renewcommand{\)}{\right)}

\def\nk         {n{\bf k}}

\def\dk         {\frac{d\,{\bf k}}{\(2 \pi\)^3}}
\def\dk1        {\frac{d\,{\bf k}_1}{\(2 \pi\)^3}}
\def\dq         {\frac{d\,{\bf q}}{\(2 \pi\)^3}}

\def\dk         {\frac{d\,{\bf k}}{\(2 \pi\)^3}}

\def\dq         {\frac{d\,{\bf q}}{\(2 \pi\)^3}}

\def\Vscf       {\hat{V}_{scf} }

\def\se         {self-energy }

\def\nk         {n{\bf k}}

\def\kmq        {{\bf k-q}}
\def\pmq        {{\bf p-q}}

\def\uu		{{\bf u}_{Is}}

\def\g15	{$\Gamma_{15}$}
\def\x1	    {$X_{1}$}
\def\l1	    {$L_{1}$}

\def\bql         {\hat{b}_{\qq \gl}}

\def\bdql        {\hat{b}^{\dg}_{-\qq \gl}}

\def\gsq         {{\mid g^{\gql}_{n' n \kk} \mid}^2}

\def\bverb      {\renewcommand{\baselinestretch}{1}\begin{verbatim}}
\def\everb  	{\end{verbatim}\renewcommand{\baselinestretch}{1.3}}

\begin{document}
\title{Competition between the electronic and phonon--mediated scattering channels in the out--of--equilibrium carrier 
dynamics of semiconductors: an {\em ab--initio} approach}

\author{Andrea Marini}

\address{Istituto di Struttura della Materia of the National Research Council, Via Salaria Km 29.3,
I-00016 Monterotondo Stazione, Italy.}

\ead{andrea.marini@cnr.it}

\begin{abstract}
The carrier dynamics in bulk Silicon, a paradigmatic indirect gap semiconductor, is studied by using 
the Baym--Kadanoff equations. Both the electron--electron\,(e--e) and electron--phonon\,(e--p) self--energies
are calculated fully \ai by using a semi--static out--of--equilibrium $GW$ approximation in the e--e case and a Fan self--energy
in the e--p case. By using the generalized Baym--Kadanoff ansatz the two--time evolution is replaced by the only dynamics
on the macroscopic time axis. 
The enormous numerical difficulties connected with a real--time simulation of realistic systems is overcome by using 
a completed collision approximation that further simplifies the memory effects connected to the time evolution.
The carrier dynamics is shown to reduce in such a way to have stringent connections to the well--known equilibrium
electron--electron and electron--phonon self--energies. 
This link allows to use general arguments to motivate the relative balance between the
e--e and e--p scattering channels on the basis of the carrier energies.
\end{abstract}

\section{Introduction}
Since the seminal works of two chemists, Norrish and Porter (recognized by the 1976 Nobel prize in Chemistry),
laser technology has grown explosively until reaching in recent years laser
durations of few attoseconds and power flux densities larger than 10$^{18}$ W/cm$^2$~\cite{Krausz2009}. 
This led to the opening of the new era of femtosecond\,(fs) nanoscale physics,
where combined non--linear and non--adiabatic phenomena~\cite{Rossi2002} occur and where it is 
possible to monitor real--time the dynamics of photo--induced electronic excitations
with unprecedented precision~\cite{Goulielmakis2010/08/05/print}. 

Ultrafast science at the nano-scale is clearing the path for nanostructure devices with efficient light emission spectra, high optical gain
and controllable atomic deformations, thus paving the way to strategic applications in chemistry, biophysics and medicine.
The design of nano-scale devices is, however, inevitably linked to a detailed knowledge of the structural and dynamical properties of these systems. Despite the
massive number of available experimental results there are still scarce numerical and theoretical methods in use of the scientific community. 

Nanostructures
and biological systems are, indeed, formed by hundreds/thousands of atoms and their peculiar properties are related to their 
reduced dimensionality and extended surface. Any
reliable theory is inevitably linked to a detailed knowledge of their structural and dynamical properties. 
Due to the complexity of these systems, however, state-of-the
art methods are confined to rather simple models with empirical parameters. In this way the theory is deprived of its predictive aspiration, and of the possibility of
inspiring new experiments and practical applications. This situation is unavoidably creating a gap between theory and experiment.

Density Functional Theory\,(DFT)\cite{R.M.Dreizler1990} represents the most up--to--date, systematic and predictive approach to 
study the electronic, structural and optical properties of an impressive wide range of materials\cite{Kohn1999}, including solids and
nanostructures.
DFT is, however, a ground--state theory that, as it will be shortly discussed in this paper, can only provide a suitable, parameter--free (and thus
predictive) basis where the actual real--time simulations are carried on. All the class of methods developed starting from the 
DFT are labeled as \ai\, methods.

It is well known that extended systems have been mostly studied by using the many-body  perturbation
theory\,(MBPT)~\cite{strinati} and Time--Dependent Density--Functional--Theory\,(TDDFT)\cite{Marques2012}.
The different treatment of correlation and nonlinear effects marks the range of applicability of the two approaches. The real-time TDDFT makes
possible to investigate nonlinear effects like second harmonic generation\cite{takimoto:154114} or hyperpolarizabilities of 
molecular systems\cite{PSSB:PSSB200642067}. However the standard approaches used to approximate the exchange-correlation functional of TDDFT 
treat correlation effects only on a mean-field level. As a consequence, while finite systems---such as molecules---are well described, 
in the case of extended systems---such as periodic crystals and nano-structures---the real-time TDDFT does not capture the essential 
features of the optical absorption~\cite{Onida2002} even qualitatively. 
On the contrary MBPT allows to include correlation effects using controllable and systematic approximations for the self-energy $\Sigma$,
that is a one-particle operator non-local in space and time.

\section{The equilibrium and out--of--equilibrium dynamics in an \ai\, framework}
\label{aiNEGF}
In recent years, the MBPT approach has been merged with DFT by using the Kohn-Sham~\cite{Marques2012} Hamiltonian as zeroth-order term in the
perturbative expansion of the interacting Green's functions. This approach is parameter free and completely \emph{ab-initio},~\cite{Onida2002} 
and in this paper will be addressed as {\it ab-initio}-MBPT ({\it Ai}-MBPT) to mark the difference with the conventional MBPT. 

In practice the Kohn--Sham equation is first solved:
\begin{align}
\[-\frac{\nabla^2}{2}+V_{KS}\[\{\RR\}\]\(\rr\)\]\phi_{n\kk}\(\rr\)=\gee_{n\kk} \phi_{n\kk}\(\rr\), 
\label{eq:ks_1}
\end{align}
with
\begin{align}
V_{KS}\[\{\RR\}\]\(\rr\)=V_{ion}\[\{\RR\}\]\(\rr\)+\intrp v\(\rr,\rr'\)n\(\rr'\)+V_{xc}\(\[n\],\rr\).
\label{eq:ks_2}
\end{align}
$V_{KS}$ is, thus, written in terms of the bare Coulomb interaction $v\(\rr_1,\rr_2\)\equiv\left|\rr_1-\rr_2\right|^{-1}$, 
the electronic density $n\(\rr\)$, the exchange--correlation potential $V_{xc}\(\[n\],\rr\)$ and the total ionic potential
$V_{ion}\[\{\RR\}\]\(\rr\)$:
\begin{align}
V_{ion}\[\{\RR\}\]\(\rr\)=\sum_{I} V_{ion}\(\rr-\RR_{I}\),
\label{eq:ks_3}
\end{align}
with $\RR_I$ the position of the generic atom $I$. All the simulations presented in this work have been carried on by using 
the Quantum Espresso\cite{Giannozzi2009} and the Yambo\cite{AndreaMarini2009} by using pseudo--potential approach written in a plane--waves basis.
In this approach $V_{ion}\(\rr-\RR_{I}\)$ is a pseudo--potential defined only for the valence electrons (8 in the case of bulk Silicon) in order to restrict the 
electronic dynamics to valence electron only. Therefore core--electrons are not considered in the present work. 

By plugging into Eq.\ref{eq:ks_3} the precise atomic structure of a given
material, the solution of Eq.\ref{eq:ks_1} provides a basis for the electronic levels suitable to rewrite the Many--Body problem. Indeed 
any Green's function $G\(1,2;\gb\)$ can be easily rewritten as a matrix in the KS basis
\begin{align}
G\(1,2;\gb\)=\sum_{nn'\kk} \[\GG_{\kk}\(t_1,t_2;\gb\)\]_{nn'} \phi^*_{n\kk}\(\rr_1\) \phi_{n'\kk}\(\rr_2\),
\label{eq:ks_3a}
\end{align}
where $1=\(\rr_1,t_1\)$ and $2=\(\rr_2,t_2\)$ represent global position, 
time and (eventually) spin indexes and $\gb=k_B T_{th}$ with $k_B$ the Boltzmann constant and 
$T_{ph}$ the equilibrium temperature.

However the {\it Ai}-MBPT is a very cumbersome technique that, based on a perturbative concept, increases its level of complexity with
the order of the expansion. As an example, this makes the extension of this approach beyond 
the linear response regime quite complex, though there have been recently some applications of the {\it Ai}-MBPT in 
nonlinear optics~\cite{Chang2002,Leitsmann2005,PhysRevB.82.235201}.

Another stringent restriction of the {\it Ai}-MBPT is that it cannot be applied when non-equilibrium phenomena take place: for 
example it cannot be applied to study the light emission after an ultrafast laser pulse excitation.
A generalization of MBPT to non-equilibrium situations has been proposed by Kadanoff and Baym\cite{Kadanoff1962,HartmutHaug2008,Schafer2002,Bonitz1998}.
In their seminal works the authors derived a set of equations for the real-time Green's functions, the Kadanoff-Baym equations (KBE's), that 
provide the basic tools of the non-equilibrium Green's Function theory and allowed essential advances 
in non equilibrium statistical mechanics.

In the equilibrium MBPT, due to the time translation invariance,
the relevant variable used to calculate the Green's functions is the frequency $\omega$. Instead, out of equilibrium, 
the time variables acquire a special role and much more attention is devoted to the their propagation properties. 
The time propagation avoids the explosive dependence, beyond the linear response, of the MBPT on high order Green's functions. Moreover the KBE's are
non-perturbative in the external field therefore weak and strong fields can be treated on the same footing. 

The key ingredient of the KBE's are the two times lesser/greater Green's functions\cite{Kadanoff1962,HartmutHaug2008,Schafer2002,Bonitz1998} 
$G^{\lessgtr}_{n_1n_2\kk}\(t_1,t_2\)$ and self--energies $\Sigma^{\lessgtr}_{n_1n_2\kk}\(t_1,t_2\)$. If we consider the time evolution of 
$G^{<}$ only on the macroscopic axis ($t_1=t_2=T$) we get the starting form of the KBE's I will further analyze in the later sections of 
this paper:
\begin{multline}
i \frac{\partial}{\partial T} \GG^<_\kk\(T,T;\gb\)=
\[\hh^{KS}_{\kk}+\VV^{Hartree}_{\kk}\[n\(T;\gb\)\]+\capo \SiS^{s}_{\kk}\(T;\gb\)+\UU_{\kk}\(T;\gb\),
\GG_{\kk}^<\(T,T;\gb\) \]+\bSS^<_{\kk}\(T;\gb\).
\label{eq:KBE_1}
\end{multline}
In Eq.\ref{eq:KBE_1} all quantities are matrices in the space of KS states (see Eq.\ref{eq:ks_3a}), e.g., $\Sigma^{s}_{n_1n_2\kk}$, $\gb=\(k_B T_{el}\)^{-1}$ where $T_{el}$
is the phonon bath temperature, and
I have already introduced the fundamental ingredients that will be discussed in the following of this paper:
\begin{itemize}
\item{} The equilibrium KS Hamiltonian, $\hh^{KS}$.
\item{} The Hartree potential $\VV^{Hartree}$. The ionic and kinetic part is already embodied in $\hh^{KS}$.
\item{} The static self--energy $\SiS^{s}$. This is responsible for introducing static e--e correlations that, if properly
defined~\cite{Attaccalite2011b} reduce the KBE, in the linear--response regime, to 
the well--known Bethe--Salpeter equation\cite{strinati}.
\item{} The interaction with the external, arbitrary electric field. The only assumption here is the the field wavelength is long compared 
with the unit cell size (optical limit).
\item{} The interaction kernel $\bSS^<$. This is the most important part of the equation that embodies dynamical e--e and e--p scatterings. It will be
defined and carefully analyzed in Sec.\ref{NEQ}.
\end{itemize}
One of the first attempts to apply the KBE's for investigating optical properties of semiconductors was 
presented in the seminal paper of Schmitt-Rink and co-workers.~\cite{PhysRevB.37.941} 
Later the KBE's were applied to study quantum wells,~\cite{PhysRevB.58.2064} laser excited
semiconductors,~\cite{PhysRevB.38.9759} and luminescence~\cite{PhysRevLett.86.2451}. However, only recently it was possible to simulate the Kadanoff-Baym
dynamics in real-time.~\cite{Kohler1999123,PhysRevLett.103.176404,PhysRevLett.84.1768,PhysRevLett.98.153004,Pal2011} 

Regarding the \ooe carrier dynamics there is a vast literature~\cite{Schafer2002} where the KBE's have been solved using several kind of
approximations. I do not want to give here an exhaustive list of all the works in this field, but a key aspect of most of them is that they rely on 
very strong approximations for what regards the underlying electronic structure and microscopic dielectric properties of the materials used. For
example a large number of works use few (two or four) bands models like in Refs.\cite{Gartner2000,Sun2012,Vu1999,Jahnke1995,Banyai1998,ElSayed1998}. In all these works the
electronic levels are approximated as well as the screening of the electron--electron interaction. 

In this work I want to go well beyond these approximations by solving the KBE's in a fully \ai framework. I will refer to this novel methodology as
\ai Non--Equilibrium Green's function\,({\em Ai}--NEGF) theory. 

\section{The equilibrium self--energies}
In order to discuss the physics involved in the dynamics of carriers taken out--of--equilibrium it is useful to
introduce the properties of the two self--energies that will be used to describe the e--e and e--p coupling. In the next two subsections, indeed,
I will introduce the approximations used and the key definitions of the different ingredients needed to build the self--energies.
I will later rewrite these equilibrium self--energies in the framework of the {\em Ai}--NEGF  theory.

The e--e interaction will be treated in the so called $GW$ approximation\cite{Aulbur19991} where the self--energy is expanded at the first order in the 
dynamically screened interaction. This approximation has provided a priceless tool to evaluate accurately the electronic levels of a wide class of
materials~\cite{Onida2002}.

Since its first application to semiconductors\cite{PhysRevLett.45.290} the $GW$ self-energy has been shown to
correctly reproduce quasi-particle energies and lifetimes for a wide range of materials\cite{Aulbur19991}.
Furthermore, by using the static limit of the $GW$ self-energy as a scattering potential of the
Bethe-Salpeter equation (BSE)\cite{strinati}, it is possible to calculate response functions including electron-hole interaction
effects.

In contrast to the e--e case the e--p coupling has received only minor attention in the \ai community. And this is somehow astonishing 
if we consider that the e--p coupling is well known to play a key role in several physical phenomena.
For example, beside inducing the well--known superconducting phase of several materials~\cite{supercond}, 
it affects the renormalization of the electronic bands\cite{allen1983,Cardona2006,marini2008,Giustino2010,cannuccia}.

Despite the development of more powerful and efficient computational resources 
the actual calculation of the effects induced by the e--p coupling in realistic materials remains
a challenging task. In addition to the numerical difficulties, it has been assumed, for a long time, that
this interaction can yield only minor corrections (of the order of meV) to the electronic levels.
As a consequence the majority of the \ai simulations of the electronic and optical properties of
a wide class of materials are generally performed by keeping the atoms frozen in their crystallographic positions. 
More importantly it is well--known that
phonons are atomic vibrations and, as a such, can be easily populated by increasing the temperature. This naive observation
is {\em de-facto} used 
to  associate the effect of the electron--phonon coupling to a temperature effect that vanishes as the temperature goes to zero. However this is not correct
as the atoms posses an intrinsic spatial indetermination due to their quantum nature, that is independent on
the temperature. 
These quantistic oscillations are taken in account by the e--p coupling when $T\rar 0$ in the form of a zero--point--motion effect.
\begin{figure}[t]
\begin{center}
\parbox[c]{7cm}{
\begin{center}
\epsfig{figure=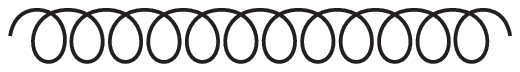,width=4cm}\\
The phonon propagator: $D_{\qq \gl}^{(0)}(\go)$
\end{center}
}
\parbox[c]{7cm}{
\begin{center}
\epsfig{figure=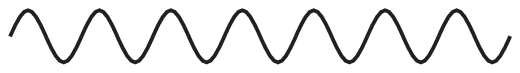,width=4cm}\\
The dynamically screened interaction $W$. 
\end{center}
}\\
\parbox[b]{7.5cm}{
\begin{center}
\epsfig{figure=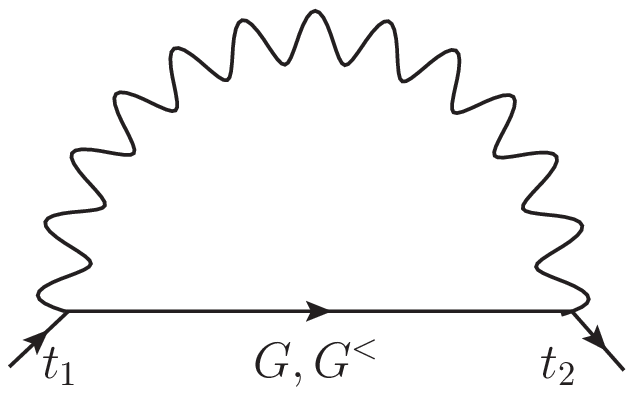,width=3cm}\\
\text{(a) The equilibrium self--energy: $\Sigma^{e-e,eq}$}
\end{center}
}
\parbox[b]{7.5cm}{
\begin{center}
\epsfig{figure=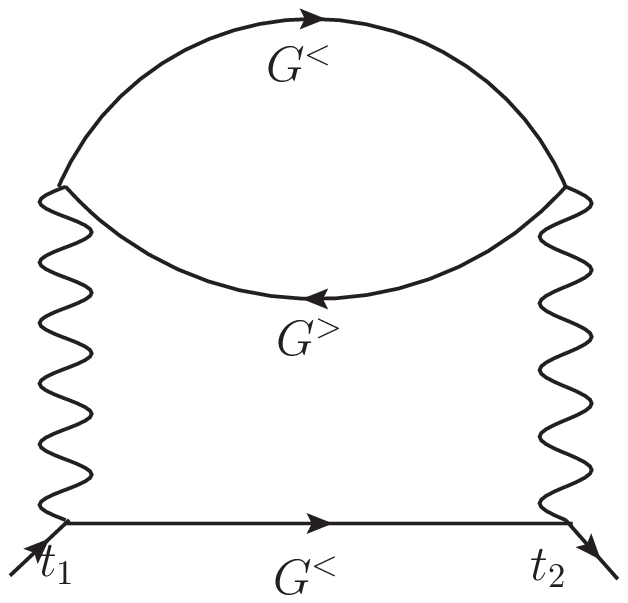,width=3cm}\\
\text{(b) The out--of--equilibrium self--energy: $\Sigma^{e-e}$}
\end{center}
}\\
\parbox[b]{7cm}{
\begin{center}
\epsfig{figure=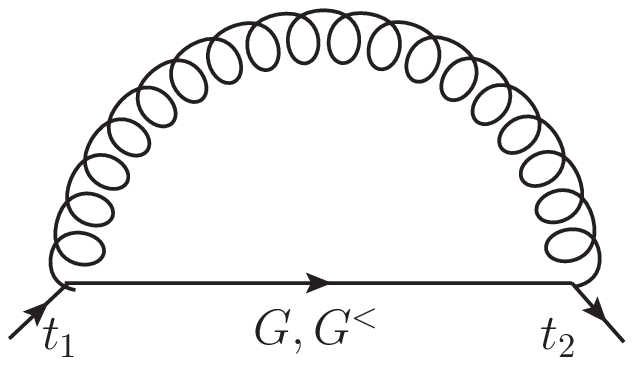,width=4cm}\\
\text{(c) The Fan self--energy: $\Sigma^{e-p}$}
\end{center}
}
\end{center}
\caption{\footnotesize{
The \se diagrams at the equilibrium and out--of--the--equilibrium used to account for e--e and e--p coupling. The kind of Green's functions
used in the diagrams are shown. Diagram (a) refers to the equilibrium case only, diagram (b) to the out--of--the--equilibrium case while
diagram (c) can be used both at and out--of--the equilibrium. This is beacause in this work the phonons are assumed to be in thermal
equilibrium with the external temperature. This means that no phonon dynamics is included.
For the e--p case I consider the first order expansion in the atomic 
displacements corresponding to the Fan self--energy (diagram (c)). In the e--e case, instead, I first consider an equilibrium
$GW$ self--energy (diagram (a)) and the corresponding out--of--equilibrium version (diagram (b)) . In the diagram (c)
the independent particle polarization corresponds to the dynamical part of the diagram. Indeed the two vertical wiggled lines
correspond to a statically screened microscopical e--e interaction (see Eq.(\ref{eq:KBE_7})). 
}}
\label{fig:diagrams_1}
\end{figure}

\subsection{The $GW$ approximation for the electronic self--energy}
The Feynman diagram corresponding to the $GW$ approximation\,(GWA) in the equilibrium is showed in the panel $(b)$ of Fig.\ref{fig:diagrams_1}. It 
corresponds to the lowest order contribution to the self--energy expressed as a Taylor expansion in the
screened interaction $W\(1,2\)$:
\begin{align}
\Sigma^{eq}_{e-e}\(1,2\)=i W\(1^+,2\)G_1\(1,2\).
\label{eq:GW_1}
\end{align}
In the present case we will use a spinless formulation where spin indexes are summed out. In Eq.(\ref{eq:GW_1})
$1=\(\rr_1,t_1\)$ and $2=\(\rr_2,t_2\)$ represent global position and time. In all diagrams showed in  Fig.\ref{fig:diagrams_1}
the time lies on the real--time axis. The out--of--equilibrium version of the GWA will be introduced in the next section. 
\begin{figure}[h]
\begin{center}
\epsfig{figure=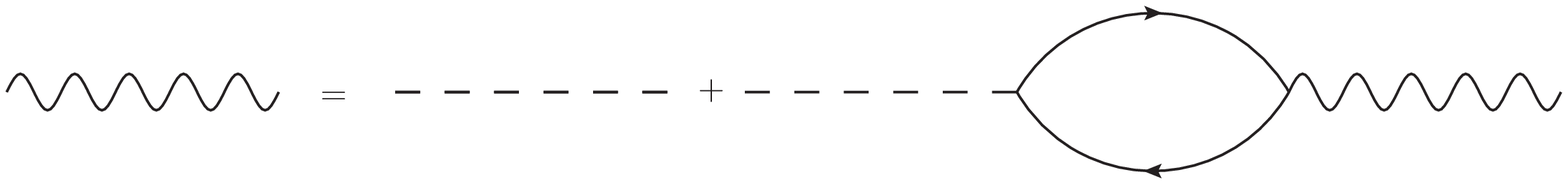,width=10cm}
\end{center}
\caption{\footnotesize{
The equation for the screened interaction $W$ within the RPA.
}}
\label{EQ_linewidths}
\end{figure}
In the GWA the screened interaction is written as:
\begin{align}
W\(1,2\)=v\(1,2\) + \int\,d34\, v\(1,3\) \tilde{\chi}\(3,4\) W\(4,2\),
\label{eq:GW_2}
\end{align}
with $v\(1,2\)\equiv v\(\rr_1,\rr_2\)\gd\(t_1-t_2\)$ is the bare Coulomb interaction and
$\tilde{\chi}$ is the irreducible electronic response function~\cite{strinati}. An approximation for $W$ follows 
by choosing $\tilde{\chi}$ to be calculated in the independent--particle approximation which corresponds, for $W$, to the 
random--phase--approximation\,(RPA), where
\begin{align}
\tilde{\chi}\(1,2\)\approx\chi_0\(1,2\)=-iG_0\(1,2\)G_0\(2,1\).
\label{eq:GW_3}
\end{align}
The actual evaluation of the space dependence appearing in Eq.\ref{eq:GW_3} is treated in reciprocal space, by summing on the lattice
vectors $\GG,\GG'$ and integrating over the Brillouin Zone\,(BZ):
\begin{align}
 \chi\(\rr_1,\rr_2;t\)=\int_{BZ}\dq \sum_{{\bf G},{\bf G}'}
 e^{i\({\bf q+G}\)\cdot {\bf r}_1} 
 \chi_{{\bf G},{\bf G}'}\({\bf q},t\)
 e^{-i\({\bf q+G}'\)\cdot{\bf r}_2}.
\label{eq:GW_4}
\end{align}
It follows that, within the Random--Phase approximation\,(RPA):
\begin{align}
 \chi_{{\bf G\;G'}}\({\bf q},\go\)=\chi^0_{{\bf G\;G'}}\({\bf q},\go\)+
 \sum_{{\bf  G''}}\chi^0_{{\bf G\;G''}}\({\bf q},\go\)
 \frac{4 \pi}{|{\bf q+G''}|^2}\chi_{{\bf G''\;G'}}\({\bf q},\go\).
 \label{eq:GW_5}
\end{align}
Eq.\ref{eq:GW_5} is used together with the definition of the single--particle Green's function
\begin{align}
 G_{0}\({\bf r}_1,{\bf r}_2;\go\)=
 2\sum_n\sum_{{\bf k}\in BZ} \phi_{n {\bf k}}\({\bf r}_1\) \phi^*_{n {\bf k}}\({\bf r}_2\) G_{\nk}\(\go\),
 \label{eq:GW_6}
\end{align}
with
\begin{align}
 G_{\nk}\(\go\)=
 \frac{f_{\nk}}{\go-\gee_{\nk}-i\gd}+\frac{1-f_{\nk}}{\go-\gee_{\nk}+i\gd};
 \label{eq:GW_7}
\end{align}
to write the final expression of the $GW$ \se. In the present work we are mainly interested in the carrier dynamics and,
in order to keep the notation as compact as possible, we restrict our attention on the imaginary part only of 
$\Sigma_{e-e}^{eq}$. This, indeed, defines the quasi--particle linewidth as
\begin{align}
\Gamma^{e-e,eq}_{\nk}=\Im\[\Sigma^{e-e,eq}_{\nk}\(\gee_{\nk}\)\],
\label{eq:GW_8}
\end{align}
where we have used the simplest on--the--mass--shell approximation\cite{Aulbur19991}. By doing simple algebraic operations it is possible to rewrite
$\Gamma^{e-e,eq}$ as:
\begin{multline}
\Gamma^{e-e,eq}_{\nk}=
\frac{2\pi}{N\Omega}\sum_{m}\sum_{\qq} \sum_{{\bf G},{\bf G}'}
\left\{\tilde{\gr}_{n m}\({\bf k},{\bf q},{\bf G}\)
  \[\tilde{\gr}_{n m}\({\bf k},{\bf q},{\bf G}'\)\]^* 
  Im\[  W_{\GG,\GG'}\(\qq,\gee_{\nk}-\gee_{m\kmq}\)\]\capo
  \[\underset{\bf electron\,scattering}{\underbrace{\(2-f_{m\kmq}\)\gt\(\gee_{\nk}-\gee_{m\kmq}\)}}-
   \underset{\bf hole\,scattering}{\underbrace{f_{m\kmq}\gt\(\gee_{m\kmq}-\gee_{\nk}\)}}\]
\right\}.
\label{eq:GW_9}
\end{multline}
The term $(electron\,scattering)$ represents an electron with energy $\gee_{\nk}$ that decays into an empty state 
with lower energy $\gee_{m\kmq}$ dissipating the corresponding energy difference in positive energy e--h pairs
(represented by $ W_{\GG,\GG'}\(\qq,\gee_{\nk}-\gee_{m\kmq}\)$). The $(hole\,scattering)$ channel is equivalent to the $(electron\,scattering)$  with the only difference  that the hole
jumps into a full state with higher energy.

In Eq.\ref{eq:GW_9} we have defined
\begin{align}
 \tilde{\gr}_{n m}\({\bf k},{\bf q},{\bf G}\)=
 \la n {\bf k}| e^{i\({\bf q+G}\)\cdot{\bf r}_1}| m \kmq \ra=
 \int\,d{\bf r}\, u^*_{\nk}\({\bf r}\) u_{m\kmq}\({\bf r}\) e^{i\({\bf q+G}\)
 \cdot{\bf r}},
 \label{eq:GW_10}
\end{align}
where $u_{\nk}\({\bf r}\)$ is the periodic part of the Bloch function. We have also introduced the dynamically screened interaction
$W_{\GG,\GG'}\(\qq,\go\)$ defined as
\begin{align}
 W_{\GG,\GG'}\(\qq,\go\)=
 \frac{\(4\pi\)}{|\qq+\GG|^2}\gd_{\GG,\GG'}+
 \frac{\(4\pi\)}{|\qq+\GG|^2}\chi_{{\bf G},{\bf G}'}\({\bf q},\go\)\frac{\(4\pi\)}{|\qq+\GG'|^2}.
 \label{eq:GW_11}
\end{align}

\subsection{The electron--phonon Fan self--energy}
The phonon contribution to the equilibrium \se comes from the oscillations of the atoms around their equilibrium positions.
The vibrational states of the system are described, fully \ai, 
by using the well--known extension of DFT, Density Functional Perturbation Theory\,(DFPT)\cite{baroni2001,Gonze1995} where the change in the 
ionic potential due to a generic atomic displacement is taken in account, self--consistently, with the solution of
the KS equation. 
In the DFPT framework the bare ionic potential is screened by the electronic polarization and the interaction Hamiltonian between
electrons and phonons $\hat{H}_{el-at}$ is given by
\begin{align}
\hat{H}_{el-at}=\int_{crystal}\,d\rr\, \hat{\gr}\(\rr\) \Vscf\[\{\RR\}\]\(\rr\).
\label{eq:FAN_1}
\end{align}
In Eq.(\ref{eq:FAN_1}) the self--consistent ionic potential, calculated at the first order in the atomic
displacements, is given by\cite{baroni2001,Gonze1995}
\begin{align}
\Vscf\[\{\RR\}\]\(\rr\)=\int\,d\rr_1\[\gd\(\rr-\rr_1\)+\int\,d\rr_2 f_{Hxc}\(\rr,\rr_2\) 
\underline{\chi}\(\rr_2,\rr_1\)\] V_{ion}\[\{\RR\}\]\(\rr_1\),
\label{eq:FAN_2}
\end{align}
where $\underline{\chi}$ is the TDDFT response function calculated using the $f_{Hxc}$ kernel. In this work I have used the Local Density
Approximation for $V_{xc}$ and $f_{Hxc}$.
If we now consider a configuration of lattice displacements ${\uu}$, $H$ can be expressed as a Taylor expansion
\begin{align}
\hat{H}_{el-at}-\overline{\hat{H}}_{el-at} = \sum_{I \ga} \overline{\frac{\partial \Vscf\[\{\RR\}\]\(\rr\)}{\partial{R_{I\ga}}}} \hat{u}_{I \ga},
\label{eq:FAN_3}
\end{align}
where $\ga$ is the Cartesian coordinate. Any overlined operator $\overline{\hat{O}}$  is evaluated with the atoms
frozen in their crystallographic equilibrium positions.
The link with the perturbative expansion is readily done by transforming Eq.\,(\ref{eq:FAN_3}) from the space of 
the lattice displacements to the space of the canonical lattice vibrations by means of the identity\cite{mattuck}: 
\begin{align}
\hat{u}_{I \ga}=\sum_{\qq \gl} \(2 N_q M_I \go_{\qq \gl}\)^{-1/2} \xi_{\ga}\(\qq \gl|I\) e^{i \qq\cdot\(\RR_I\)}
\(\hat{b}^{\dagger}_{-\qq \gl}+\hat{b}_{\qq \gl}\),
\label{eq:FAN_4}
\end{align}
where $N$ is the number of cells (or, equivalently the number of q--points) used in the simulation and $M_I$ is the atomic mass
of the $I$--th atom in the unit cell. $\xi_{\ga}\(\qq \gl|I\)$ is the phonon polarization vector
and $\bdql$ and $\bql$ are the bosonic creation and annihilation operators.

By inserting Eq.\,(\ref{eq:FAN_4}) into Eq.\,(\ref{eq:FAN_3}) we get
\begin{align}
\hat{H}-\overline{\hat{H}} 
= \frac{1}{\sqrt{N_q}}\sum_{\kk n n^{'} \qq \gl} g^{\gql}_{n' n \kk} 
c^{\dagger}_{n'\kk+\qq} c_{n\kk} \(\hat{b}^{\dagger}_{-\qq \gl} +\hat{b}_{\qq \gl} \)
\label{eq:FAN_5}
\end{align}
In Eq.\,(\ref{eq:FAN_5}) we have introduced the  ($g^{\gql}_{n' n \kk}$) 
electron--phonon matrix elements. This is obtained by rewriting $V_{scf}$ and by making explicit its dependence on the
atomic positions: 
\begin{align}
\Vscf\[\{\RR\}\]\(\rr\)=\sum_{Is} \Vscf^{\(s\)}\(\rr-\RR_{Is}\).
\label{eq:FAN_6}
\end{align}
By using Eq.\ref{eq:FAN_6} I get an explicit form for the 
electron--phonon matrix elements
\begin{align}
g^{\gql}_{n' n \kk}=\sum_{s \ga} \(2 M_s \go_{\gql}\)^{-1/2} e^{i\qq\cdot\tau_s} 
\la n'\kk+\qq |\frac{\partial \Vscf^{\(s\)}\(\rr\)}{\partial{R_{s\ga}}} | n \kk \ra \xi_{\ga}\(\qq \gl|s\).
\label{eq:FAN_7}
\end{align}
We have used the short 
form $R_{s \ga}=\left. R_{Is\ga}\right|_{I=0}$.

Note that, in principle, the interaction Hamiltonian, Eq.(\ref{eq:FAN_5}), can be also used to obtain a diagrammatic 
representation of the phonon self--energy. If the electronic and phononic problems is solved at the same time, however,
the exchange--correlation effects introduced within the DFPT would lead to severe double counting problems~\cite{RvL2004}.
In the present case this problem is avoided by using bare DFPT phonon modes without taking in account additional 
Many--Body corrections.

The actual calculation of the Fan diagram is straightforward. The Fan diagram is similar to the $GW$ one 
where the screened electronic interaction $W$ is replaced by a phonon 
propagator of wavevector $\qq$ and branch $\gl$\cite{mahan}.
By applying the finite temperature diagrammatic rules it is possible to 
define the Fan \se operator $\Sigma^{e-p}_{n\kk}(\go)$, recovering the expression originally evaluated by Fan\cite{fan1950} for the phonon mediated
quasi--particle linewidth
\begin{align}
\Gamma^{e-p,eq}_{\nk}\(\gb\)=\Im\[\Sigma^{e-p,eq}_{\nk}\(\gee_{\nk},\gb\)\],
\label{eq:FAN_8}
\end{align}
with
\begin{multline}
\Gamma^{e-p,eq}_{n\kk}(\gb) =
\left.\Gamma^{e-p,eq}_{n\kk}\rb_{e}+
\left.\Gamma^{e-p,eq}_{n\kk}\rb_{h}+\left.\Gamma^{e-p,eq}_{n\kk}\(\gb\)\rb_{ph} =\\
\frac{\pi}{N}\sum_{n'\gql} \gsq  \(1-f_{n'\kk-\qq}\)\gd\(\gee_{n'\kmq}-\gee_{n\kk} +\goql\)+\\
\frac{\pi}{N}\sum_{n'\gql} \gsq   f_{n'\kk-\qq}  \gd\(\gee_{n'\kmq}-\gee_{n\kk} -\goql\)+\\
\frac{\pi}{N}\sum_{n'\gql} \gsq N_{\qq\gl}\(T\)\[\gd\(\gee_{n'\kmq}-\gee_{n\kk} +\goql\) +  \gd\(\gee_{n'\kmq}-\gee_{n\kk} -\goql\)\],
\label{eq:FAN_9}
\end{multline}
with $N_{\qq\gl}\(T\)$ is the Bose function distribution of the phonon mode $\(\qq,\gl\)$ at temperature $T$.
The e--p lifetime is splitted in Eq.\ref{eq:FAN_9} in three different contributions corresponding to the 
scattering of an electron/hole $\left.\Gamma^{e-p,eq}_{n\kk}\rb_{e/h}$ with the phonon bath and an explicitly temperature 
term, $\left.\Gamma^{e-p,eq}_{n\kk}\(\gb\)\rb_{ph}$.
The Fan \se can be also derived by using the framework of the
Heine--Allen--Cardona\,(HAC) theory~\cite{allen1976,allen1983,Cardona2006}. The HAC approach is based on the 
static Rayleigh-Schr\"{o}dinger perturbation theory done using the $\hat{u}_{Is\ga}$ displacement operators
as scalar variables.

\subsection{Electron--phonon versus electron--electron contribution to the equilibrium electronic lifetimes}
\begin{figure}[t]
\begin{center}
\epsfig{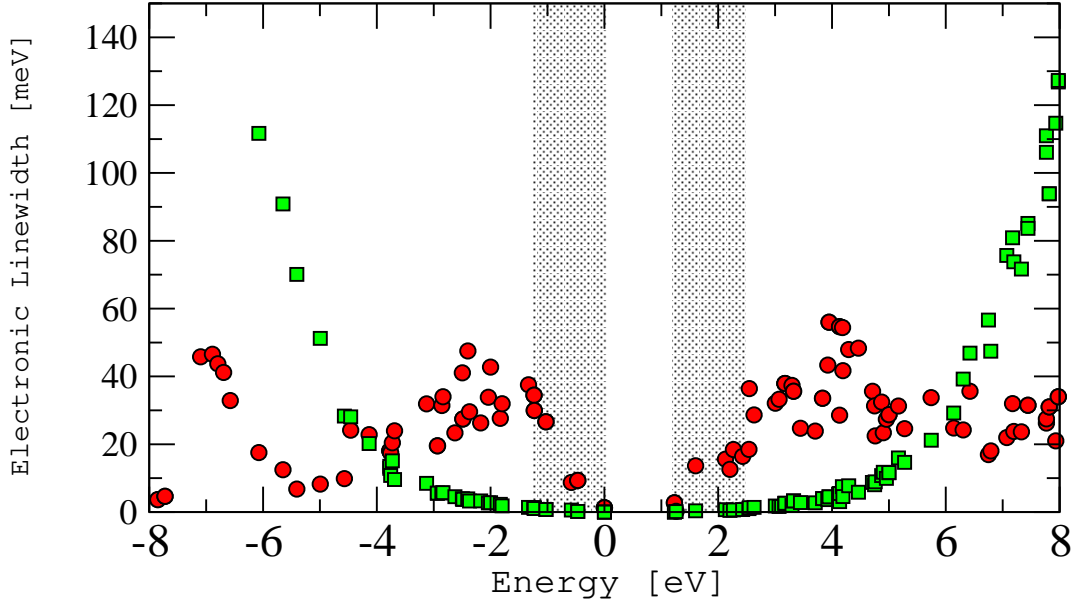}
\end{center}
\caption{\footnotesize{
Quasiparticle linewidths of bulk Silicon calculated by using the $GW$ approximation for 
the e--e scattering (green boxes) and the Fan approximation for the e--p scattering (red circles). The two gray
areas denote the energy regions (as large as twice the electronic indirect gap $2E_g\approx$1.2\,eV) 
where, by simple energy conservation arguments, the electronic linewidth are 
zero by definition. In these energy conditions the e--p contribution is stronger and the corresponding linewidths are larger then the 
e--e ones. The quadratic energy dependence of the e--e linewidths inverts this trend at about five times the 
indirect gap.
}}
\label{fig:EQ_linewidths}
\end{figure}
From Eq.\ref{eq:GW_9} and Eq.\ref{eq:FAN_9} we can deduce a simple physical property of the equilibrium self--energies. By simple
energetic conditions it is clear that in a system with a direct or indirect gap $E_g$ the e--e linewidths $\Gamma^{e-e,eq}_{\nk}$ are 
exactly zero whenever $\gee_{\nk} < \gee_{CBM}+2 E_g$ or $\gee_{\nk} > \gee_{VBM}-2 E_g$, with CBM referring to the condunction band minumum and
VBM to the valence band maximum. These two energy conditions define the energy ranges
represented by gray areas in Fig.\ref{fig:EQ_linewidths}.

On the contrary the e--p linewidths, thanks to the continuum of phonon frequencies, have no zero energy regions. It follows that
in the  case  of low energy electrons or hole the e--p linewidths largely dominate on the e--e part. Only above (or below) the energy forbidden regions the e--e
linewidths start growing quadratically, following the Fermi liquid expected behavior, and cross the e--p part when $\gee_{\nk}=-4$\,eV (valence bands) and
$6$\,eV.

\section{An out--of--equilibrium formulation of the carrier dynamics including e--e and e--p scatterings}
\label{NEQ}
Now that the equilibrium self--energies have been introduced and defined we can move in the \ooe regime by giving a precise
form to the interaction kernel $\bSS_{\kk}\(T\)$. First of all this is formally defined~\cite{Kadanoff1962,HartmutHaug2008,Schafer2002,Bonitz1998} as
\begin{multline}
\bSS^<_{\kk}\(T;\gb\)=\int_{-\infty}^{T}d\gu\,
\[\SiS^{>}_{\kk}\(T,\gu;\gb\)\GG^{<}_{\kk}\(\gu,T;\gb\)+
\GG^{<}_{\kk}\(T,\gu;\gb\)\SiS^{>}_{\kk}\(\gu,T;\gb\)-\capo
\SiS^{<}_{\kk}\(T,\gu;\gb\)\GG^{>}_{\kk}\(\gu,T;\gb\)-
\GG^{>}_{\kk}\(T,\gu;\gb\)\SiS^{<}_{\kk}\(\gu,T;\gb\)\].
\label{eq:KBE_2}
\end{multline}
The key point in Eq.\ref{eq:KBE_2} is that the definition of $\bSS$ is not closed as it involves two--times functions and both the 
greater and lesser Green's functions. This means that one should consider the two--time evolution of the KBE together with the equation of 
motion for the $\GG_{\kk}^{\lessgtr}$ 
and, as we will shortly see, for the $\GG_{\kk}^{\(r/a\)}$  Green's functions.

As a two--time evolution of the KBE is, at the present stage, computationally not feasible for the real--time dynamics of complex solids 
and nanostructures we want to explore a different path based on several key ansatz and approximations. 

For the e--e case the extension of the GWA, Eq.\ref{eq:GW_2}, to the \ooe regime is well--known~\cite{HartmutHaug2008}. The diagrammatic form 
is represented in the frame (b) of Fig.\ref{fig:diagrams_1}. What I want to do here is the rewrite this
extension in a \ai framework introducing some key approximations.  First of all I split $\SiS$ and $\bSS$ in the e--e and e--p parts:
\begin{align}
 \SiS^{\lessgtr}_{\kk}\(t_1,t_2;\gb\)=\SiS^{\(e-e\)\lessgtr}_{\kk}\(t_1,t_2;\gb\)+\SiS^{\(e-p\)\lessgtr}_{\kk}\(t_1,t_2;\gb\),
\label{eq:KBE_3}
\end{align}
\begin{align}
 \bSS^{\lessgtr}_{\kk}\(T;\gb\)=\bSS^{\(e-e\)\lessgtr}_{\kk}\(T;\gb\)+\bSS^{\(e-p\)\lessgtr}_{\kk}\(T;\gb\).
\label{eq:KBE_3a}
\end{align}
The GWA for $\SiS^{\(e-e\)\lessgtr}$ \ooe reads:
\begin{multline}
 \Sigma^{\(e-e\)\lessgtr}_{n_1n_2\kk}\(t_1,t_2;\gb\)=
  \frac{-i}{2 N\Omega}\sum_{m_1,m_2}\sum_{\GG_1,\GG_2}\sum_{\qq}
  \{\tilde{\gr}_{n_1 m_1}\({\bf k},{\bf q},{\bf G}_1\)
  \[\tilde{\gr}_{n_2 m_2}\({\bf k},{\bf q},{\bf G}_2\)\]^*\times \\
  W^{\lessgtr}_{\GG_1\GG_2}\(\qq,t_1,t_2;\gb\)
  G^{\lessgtr}_{m_1 m_2\kmq}\(t_1,t_2;\gb\)\},
\label{eq:KBE_4}
\end{multline}
with the screened interaction $W^{\lessgtr}$ written as
\begin{multline}
 W^{\lessgtr}_{\GG_1,\GG_2}\(\qq,t_1,t_2\)= 
 \int_{-\infty}^{t_1}d\gu_1
 \int_{-\infty}^{t_2}d\gu_2
 \sum_{\GG_3,\GG_4}
 W^{\(r\)}_{\GG_1,\GG_3}\(\qq,t_1,\gu_1;\gb\)\times\\
 L^{\lessgtr}_{\GG_3,\GG_4}\(\qq,\gu_1,\gu_2;\gb\)
 W^{\(a\)}_{\GG_4,\GG_2}\(\qq,\gu_2,t_2;\gb\).
\label{eq:KBE_5}
\end{multline}
In Eq.\ref{eq:KBE_5} $W^{\(r/a\)}$ are the retarded/advanced screened interactions and $L^{\lessgtr}$ is the independent particle polarization
diagram represented by the closed loop in the diagram (b) of Fig.\ref{fig:diagrams_1}:
\begin{multline}
 L^{\lessgtr}_{\GG_1,\GG_2}\(\qq,t_1,t_2;\gb\)= \frac{-i}{2N\Omega}\sum_{m_2,m_3}\sum_{m'_2,m'_3}
  \[\tilde{\gr}_{m_3 m'_2}\(\pp,\qq,\GG_1\)\]^* 
    \tilde{\gr}_{m'_3 m_2}\(\pp,\qq,\GG_2\)\times\\
  G^{\lessgtr}_{m_3 m'_3\pp}\(t_1,t_2;\gb\)
  G^{\gtrless}_{m'_2 m_2\pmq}\(t_2,t_1;\gb\).
\label{eq:KBE_5a}
\end{multline}
Before analyzing in detail the temporal dependence of $\SiS^{\(e-e\)\lessgtr}$ it is instructive to introduce the \ooe e--p self--energy.
The extension \ooe of the FAN self--energy is straightforward\cite{HartmutHaug2008}:
\begin{align}
 \Sigma^{\(e-p\)\lessgtr}_{n_1n_2\kk}\(t_1,t_2;\gb\)=
  \frac{-i}{N}\sum_{m_1,m_2}\sum_{\qq,\gl}
  \(g^{\gql}_{n_1 m_1 \kk}\)^*
  g^{\gql}_{n_2 m_2 \kk}
  D^{\lessgtr}_{\gl}\(\qq,t_1,t_2;\gb\)
  G^{\lessgtr}_{m_1 m_2\kmq}\(t_1,t_2;\gb\)\},
\label{eq:KBE_6}
\end{align}
with $D_{\gl}$ the phonon propagator that I will shortly define.

Now I introduce the complete set of approximations I will use to transform Eq.\ref{eq:KBE_4} in a simpler set of equations of motion to describe, \ai, the carrier
dynamics in realistic materials. These approximation are guided by simple physical arguments introduced with the aim of limiting the computation
effort. This is a crucial requirement to make the simulations feasible. At the same time those approximations represent a first
step in the device of a sound and complete framework for the {\em Ai}--NEGF approach and their validity will be carefully analyzed in the further 
development of this field. 

The approximations are:
\begin{itemize}
\item[(i)] The screened interaction entering Eq.\ref{eq:KBE_5} are assumed to be static. I introduce this semi--static approximation\,(SSA) of the \ooe version of
the GWA being inspired by the well--known Coulomb-Hole plus Screened-Exchange approximation\,(COHSEX)\cite{Onida2002}. The SSA reads:
\begin{align}
W^{\(r/a\)}_{\GG_1,\GG_2}\(\qq,t_1,t_2;\gb\)\approx W_{\GG_1,\GG_2}\(\qq\).
\label{eq:KBE_7}
\end{align}
This approximation is well motivated in semiconductors and insulators as an electronic gap of 1\,eV corresponds to a time scale of $0.66$\,fs. This a
very short time compared to the time scale of the simulations presented in the present work.
\item[(ii)] The simulation is done at a fixed temperature $\gb$ that is assumed to not change during the simulation. This approximation corresponds to 
take the phonon population at the equilibrium and assuming that their energy distribution does never deviate from a Bose distribution. In practice
this means that:
\begin{align}
D^{<}_{\gl}\(\qq,t_1,t_2;\gb\)\approx \(-i\) \sum_{\pm} N^{\pm}_{\qq\gl}\(\gb\) e^{\pm i \w_{\qq\gl}\(t_1-t_2\)}.
\label{eq:KBE_8}
\end{align}
Note that $D^{>}_{\gl}\(\qq,t_1,t_2;\gb\)=-\[D^{<}_{\gl}\(\qq,t_1,t_2;\gb\)\]^*$.
\item[(iii)] I use the Generalized Baym--Kadanoff ansatz\,(GBKA)~\cite{HartmutHaug2008,Schafer2002,Bonitz1998}.
\item[(iv)] I neglect the off--diagonal matrix elements of the $\GG_{\kk}^{\lessgtr}$ and $\SiS_{\kk}^{\lessgtr}$ matrices
\begin{gather}
G^{<}_{nn'\kk}\(T,T;\gb\)\approx i \gd_{n n'} f_{n\kk}\(T;\gb\),
\label{eq:KBE_9}\\
G^{>}_{nn'\kk}\(T,T;\gb\)\approx i \gd_{n n'} \[f_{n\kk}\(T;\gb\)-1\],
\tag{\ref{eq:KBE_9}$'$}\\
\Sigma^{\lessgtr}_{nn'\kk}\(T,T;\gb\)\approx \gd_{n n'} \Sigma_{n\kk}\(T;\gb\)
\tag{\ref{eq:KBE_9}$''$}.
\end{gather}
\end{itemize}
As it will be showed in the following these approximations will make possible to perform the actual simulations of complex materials at 
a reasonable computational cost. 

\subsection{The collision operator in the Generalized Baym--Kadanoff ansatz}
The GBKA provides the most important tool to disentangle the complex two--times evolution by
rewriting:
\begin{align}
G^{\lessgtr}_{n\kk}\(t_1,t_2;\gb\)\approx \(i\)
\[G^{(r)}_{n\kk}\(t_1-t_2;\gb\)G^{\lessgtr}_{n\kk}\(t_2,t_2;\gb\)-G^{\lessgtr}_{n\kk}\(t_1,t_1;\gb\)G^{(a)}_{n\kk}\(t_1-t_2;\gb\)\],
\label{eq:KBE_10}
\end{align}
that, in the lesser case, acquires a simple form written in terms of the time--dependent electronic occupations:
\begin{align}
G^{<}_{n\kk}\(t_1,t_2;\gb\)\approx 
-\[G^{(r)}_{n\kk}\(t_1-t_2;\gb\)f_{n\kk}\(t_2;\gb\)-f_{n\kk}\(t_1;\gb\)G^{(a)}_{n\kk}\(t_1-t_2;\gb\)\].
\label{eq:KBE_10a}
\end{align}
A similar identity holds for the greater Green's function.
By using Eq.\ref{eq:KBE_10} the KBE's for the electronic occupations $f_{n\kk}\(T,\gb\)$ acquire a closed form  if a given analytic
form for the advanced/retarded Green's functions is given. But before discussing the specific form of the 
$(r/a)$ Green's functions I use Eq.\ref{eq:KBE_10}  to rewrite the two contributions to $\bSS$.

The algebraic passages to rewrite $\bSS^{e-p,<}$ by using the GBKA can be found, for example, in Ref.\cite{Haug1992} in the simpler case
of a Jellium model with e--p interaction treated using a Fr\"ohlich interaction $g^{\qq\gl}_{nn\kk}\approx g_{|\qq|}$ and a single optical phonon
mode.

After some involved but conceptually simple algebra it can be shown that
\begin{align}
S_{n\kk}^{e-p,<}\(T;\gb\)=N^{-1}\sum_{\qq\gl}\sum_{s=\pm}e^{is\w_{\qq\gl}T}\Biggl\{ S_{n\kk}^{\qq\gl s}\(T;\gb\) +\[ S_{n\kk}^{\qq\gl-s}\(T;\gb\)
\]^*\Biggr\},
\label{eq:KBE_11}
\end{align}
with
\begin{multline}
S_{n\kk}^{\qq\gl s}\(T;\gb\)=\(-i\)\int_{-\infty}^{T}d\gu\, e^{-is\w_{\qq\gl}\gu} \sum_{m}
|g^{\qq\gl}_{nm\kk}|^2
G^{\(r\)}_{m \kmq}\(T-\gu;\gb\)\times\\
\[ N_{\qq\gl}^{-s}\(\gb\)\(2-f_{m\kmq}\(\gu;\gb\)\)f_{n\kk}\(\gu;\gb\)-
   N_{\qq\gl}^{s}\(\gb\)f_{m\kmq}\(\gu;\gb\)\(2-f_{n\kk}\(\gu;\gb\)\)
\]\times\\
\[G^{\(a\)}_{n \kk}\(\gu;\gb\)\]^*.
\label{eq:KBE_12}
\end{multline}
As it will be clear in the next section the sum of $S_{n\kk}^{\qq\gl s}$ and of its adjoin in Eq.\ref{eq:KBE_11} is crucial to keep the
time--evolution unitary in such a way that the number of electrons is conserved. The actual time decay of the photo--excited carriers will, then,
be connected to the actual memory of the system linked with the $\gu$ integral appearing in Eq.\ref{eq:KBE_11}  rather that with the imaginary
part of the self--energy, as commonly done in equilibrium theories. This detail marks the basic physical difference between the equilibrium
QP linewidths and the \ooe QP decay time, as it will be clear from the discussion in Sec.\ref{sec:silicon}.

In the case of the e--e scattering the application of the GBKA is more involved as it must be applied to three Green's functions instead of one, like
in the e--p case. Nevertheless, once a simple form for $G^{\(r/a\)}$ will be given the physical interpretation of the collision operators will be
equally simple. By following a procedure similar to the e--p case we find that
\begin{multline}
S_{n\kk}^{e-e,<}\(T;\gb\)=\frac{1}{2\(N\Omega\)^2}\sum_{\qq \pp}\sum_{m_1 m_2 m_3}
\left|W^{\qq}_{\substack{n m_1 \kk \\ m_3 m_2 \pp }}\right|^2
\int_{-\infty}^{T}d\gu\biggl\{\\
\[
G^{\(r\)}_{m_1 \kmq}\(T-\gu;\gb\)
G^{\(r\)}_{m_3 \pp}\(T-\gu;\gb\)
G^{\(a\)}_{m_2 \pmq}\(\gu-T;\gb\)
G^{\(a\)}_{n \kk}\(\gu-T;\gb\)+\capo
G^{\(a\)}_{m_1 \kmq}\(\gu-T;\gb\)
G^{\(a\)}_{m_3 \pp}\(\gu-T;\gb\)
G^{\(r\)}_{m_2 \pmq}\(T-\gu;\gb\)
G^{\(r\)}_{n \kk}\(T-\gu;\gb\)
\]\times\\
\[ 
\(2-f_{m_1\kmq}\(\gu;\gb\)\)
\(2-f_{m_3\pp}\(\gu;\gb\)\)
f_{m_2\pmq}\(\gu;\gb\)
f_{n\kk}\(\gu;\gb\)-\capo
f_{m_1\kmq}\(\gu;\gb\)
f_{m_3\pp}\(\gu;\gb\)
\(2-f_{m_2\pmq}\(\gu;\gb\)\)
\(2-f_{n\kk}\(\gu;\gb\)\)
\]
\biggr\}.
\label{eq:KBE_13}
\end{multline}
Eq.\ref{eq:KBE_11}--\ref{eq:KBE_13} represent the expressions for the time--dependent kernel $\bSS$ that I will further simplify in the next
section. By using this expression for $\bSS$ the final equation of motion for the electronic occupation reads:
\begin{align}
 \frac{\partial}{\partial T} f_{\nk}\(T;\gb\)=
 \left.\frac{\partial}{\partial T} f_{\nk}\(T;\gb\)\right|_{coh}-S^<_{n\kk}\(T;\gb\).
\label{eq:KBE_14}
\end{align}
with
\begin{align}
\left.\frac{\partial}{\partial T} f_{\nk}\(T;\gb\)\right|_{coh}=
-\[ \hh^{KS}_{\kk}+\VV^{Hartree}_{\kk}\[n\(T;\gb\)\]+\SiS^{s}_{\kk}\(T;\gb\)+
\UU_{\kk}\(T;\gb\),\GG_{\kk}^<\(T,T;\gb\) \]_{nn}
\tag{\ref{eq:KBE_14}$'$}.
\end{align}

\subsection{The Completed Collision Approximation}
Thanks to the GKBA the KBE's can be rewritten as a set of non--linear differential equations for the occupation functions. Nevertheless the whole 
history of these functions should be integrated numerically in Eqs.\ref{eq:KBE_12}--\ref{eq:KBE_13}. In order to further simplify the 
theory I introduce two more approximations to the list 
presented in Sec.\ref{aiNEGF}:
\begin{itemize}
\item[(v)] The $G^{\(r/a\)}$ functions are assumed to have a QP form (exponential):
\begin{align}
G^{\(r\)}_{n\kk}\(T;\gb\)\approx -i e^{-iE_{\nk}\(\gb\)T -\Gamma_nk\(\gb\)T},
\label{eq:CCA_1}
\end{align}
with $G^{\(a\)}_{n\kk}\(T;\gb\)=\[G^{\(r\)}_{n\kk}\(-T;\gb\)\]^*$.
This means that we neglect any deviation from the pure exponential behavior induced by the short--time dynamics\cite{springerlink:10.1007/s100510050770} and 
the e--e and e--p correlations\cite{Haug1996a}. In eq.\ref{eq:CCA_1} the QP energies $E_{n\kk}$ and widths $\Gamma_{n\kk}$ are calculated \ai by means of 
the equilibrium GWA and Fan self--energies. The explicit expressions for  $\Gamma_{n\kk}$ are given by Eq.\ref{eq:GW_9} and
Eq.\ref{eq:FAN_9}.
\item[(vi)] I adopt the so called Completed Collision Approximation\,(CCA)\cite{Haug1992} that makes possible to solve the time--integration
appearing in Eqs.\ref{eq:KBE_12}--\ref{eq:KBE_13} analytically by assuming that the occupation functions vary only slowly on the time scale
defined by $G^{\(r\/a\)}_{\nk}\(T-\gu;\gb\)$. This time scale is given by 
$\Gamma_{n\kk}\(\gb\)^{-1}$ and, as shown in Fig.\ref{fig:EQ_linewidths}, it is of the order of $10-20$\,fs.
Under this assumption the $f$ functions can be taken out of the integrals appearing 
in Eqs.\ref{eq:KBE_12}--\ref{eq:KBE_13} and the rest of the  integral can be done analytically by using Eq.\ref{eq:CCA_1}.
\end{itemize}
Thanks to the CCA Eq.\ref{eq:KBE_12} can be rewritten as:
\begin{multline}
S_{n\kk}^{\qq\gl s}\(T;\gb\)\approx 
\(-i\)\sum_{m}
|g^{\qq\gl}_{nm\kk}|^2
\[ N_{\qq\gl}^{-s}\(\gb\)\(2-f_{m\kmq}\(T;\gb\)\)f_{n\kk}\(T;\gb\)-\capo
   N_{\qq\gl}^{s}\(\gb\)f_{m\kmq}\(T;\gb\)\(2-f_{n\kk}\(T;\gb\)\)\]\times\\
\Biggl\{
\[\int_{-\infty}^{T}d\gu\, e^{-is\w_{\qq\gl}\gu} 
G^{\(r\)}_{m \kmq}\(T-\gu;\gb\)
\[G^{\(a\)}_{n \kk}\(\gu;\gb\)\]^*\]
\Biggr\}.
\label{eq:CCA_2}
\end{multline}
By using Eq.\ref{eq:CCA_1} the $\gu$ integral reads:
\begin{align}
\int_{-\infty}^{T}d\gu\, e^{-is\w_{\qq\gl}\gu} 
G^{\(r\)}_{m \kmq}\(T-\gu;\gb\)\[G^{\(a\)}_{n \kk}\(\gu;\gb\)\]^*=
\frac{e^{-is\w_{\qq\gl}T}}{E_{m\kmq}\(\gb\)-E^*_{n\kk}\(\gb\) -i s \go_{\qq\gl}}
\label{eq:CCA_3}.
\end{align}
Thanks to Eq.\ref{eq:CCA_3} the $S_{n\kk}^{\qq\gl s}\(T;\gb\)$ functions can be rewritten in terms of three simple lifetimes: $\gc_{n\kk}^{\(e-p,h\)}\(T;\gb\)$,
$\gc_{n\kk}^{\(e-p,e\)}\(T;\gb\)$ and $\gc_{n\kk}^{\(e-p,ph\)}\(T;\gb\)$:
\begin{align}
\left. S_{n\kk}^{\qq\gl
s}\(T;\gb\)\right|_{CCA}=-\gc_{n\kk}^{\(e-p,h\)}\(T;\gb\)\[1-f_{n\kk}\(T;\gb\)\]+\gc_{n\kk}^{\(e-p,e\)}\(T;\gb\)f_{n\kk}\(T;\gb\)+\gc_{n\kk}^{\(e-p,ph\)}\(T;\gb\).
\label{eq:CCA_4}
\end{align}
In order to define and interpret the physical properties of these lifetimes I introduce two functions: $P^{abs}$ and $P^{emit}$.
\begin{align}
P^{\(abs\)}_{nn'\kk\qq}\(\gb\)=\frac{\lb\Im E_{n\kk}\(\gb\)\rb+\lb\Im E_{n\kmq}\(\gb\)\rb}{\[\Re\(E_{n\kmq}\(\gb\)-E_{n\kk}\(\gb\)\)-\go_{\qq\gl}\]^2 + 
\[\lb\Im E_{n\kk}\(\gb\)\rb+\lb \Im E_{n\kmq}\(\gb\)\rb\]^2},
\label{eq:CCA_5}
\end{align}
and
\begin{align}
P^{\(emit\)}_{nn'\kk\qq}\(\gb\)=\frac{\lb\Im E_{n\kk}\(\gb\)\rb+\lb\Im E_{n\kmq}\(\gb\)\rb}{\[\Re\(E_{n\kmq}\(\gb\)-E_{n\kk}\(\gb\)\)+\go_{\qq\gl}\]^2
+ \[\lb\Im E_{n\kk}\(\gb\)\rb+\lb \Im E_{n\kmq}\(\gb\)\rb\]^2}.
\label{eq:CCA_6}
\end{align}
These two functions represent the energy constrain in the emission ($P^{emit}$) or absorption ($P^{abs}$) of a phonon smeared 
out due to the QP approximation for the retarded Green's function that. This, indeed, 
yields a Lorentzian indetermination in the energy conservation. By using these two simple functions the three e--p
induced lifetimes can be easily rewritten as
\begin{gather}
\gc_{n\kk}^{\(e-p,h\)}\(T;\gb\)= \sum_{n' \qq \gl} \frac{\lb g^{\qq\gl}_{nn'\kk} \rb^2 }{N} P^{\(abs\)}_{nn'\kk\qq}\(\gb\) f_{n'\kmq}\(T;\gb\),
\label{eq:CCA_7}\\
\gc_{n\kk}^{\(e-p,e\)}\(T;\gb\)= \sum_{n' \qq \gl} \frac{\lb g^{\qq\gl}_{nn'\kk} \rb^2 }{N} P^{\(emit\)}_{nn'\kk\qq}\(\gb\) \(2-f_{n'\kmq}\(T;\gb\)\),
\tag{\ref{eq:CCA_7}$'$}\\
\gc_{n\kk}^{\(e-p,ph\)}\(T;\gb\)= \sum_{n' \qq \gl} \frac{\lb g^{\qq\gl}_{nn'\kk} \rb^2 }{N} N_{\qq\gl}\(\gb\)\[f_{n\kk}\(T;\gb\)-f_{n'\kmq}\(T;\gb\)\]
\[P^{\(abs\)}_{nn'\kk\qq}\(\gb\) + P^{\(emit\)}_{nn'\kk\qq}\(\gb\)\].
\tag{\ref{eq:CCA_7}$''$}
\end{gather}
The dynamics induced by the e--p kernel on the electronic occupations can be easily visualized by using Eq.\ref{eq:CCA_4} to simplify 
Eq.\ref{eq:KBE_1} for the diagonal matrix elements of $G^{<}$. If I do not consider the coherent part of the KBE equation I get
\begin{align}
\frac{\partial}{\partial T} f_{n\kk}\(\gb;T\)=\gc_{n\kk}^{\(e-p,h\)}\(T;\gb\)\[2-f_{n\kk}\(T;\gb\)\]-
\gc_{n\kk}^{\(e-p,e\)}\(T;\gb\)f_{n\kk}\(T;\gb\)-\gc_{n\kk}^{\(e-p,ph\)}\(T;\gb\).
\label{eq:CCA_8}
\end{align}
Eq.\ref{eq:CCA_8} makes the interpretation of the carrier dynamics induced by the e--p coupling straightforward. We start indeed noticing that,
simply because of their definition, $\gc_{n\kk}^{\(e-p,h\)}$ and $\gc_{n\kk}^{\(e-p,e\)}$ are always positive. In contrast 
$\gc_{n\kk}^{\(e-p,ph\)}$ can be positive or negative.

To see why this property is important let's consider two limiting cases when $f_{n\kk}\(\gb;T=0\)=0$ or $f_{n\kk}\(\gb;T=0\)=1$. In the first case,
from Eq.\ref{eq:CCA_8} it follows that, if we consider the zero temperature case,
\begin{gather}
f_{n\kk}\(\gb\rightarrow\infty;T=0\)=0\Rightarrow \frac{\partial}{\partial T} f_{n\kk}\(\gb\rightarrow\infty;T=0\)>0
\label{eq:CCA_9},\\
f_{n\kk}\(\gb\rightarrow\infty;T=0\)=2\Rightarrow \frac{\partial}{\partial T} f_{n\kk}\(\gb\rightarrow\infty;T=0\)<0.
\tag{\ref{eq:CCA_9}$'$}
\end{gather}
This simply means that an initially completely filled level will be emptied by the e--p interaction while an initially empty level will be filled. The
$\gc_{n\kk}^{\(e-p,ph\)}$ is non zero only at finite temperature and works to build up a finite temperature 
quasi--equilibrium among the carrier occupations. 
The $\gc_{n\kk}^{\(e-p,e/h\)}$ lifetimes represent a single scattering of the state 
$|n\kk\ra$ to the state $|n\kmq\ra$ with emission/absorption of a phonon mode $|\qq \gl\ra$. This process is similar to the one studied in the
equilibrium case  and, indeed, if we consider the Lorentzian function appearing in $P^{abs}$ and $P^{emit}$ as approximated delta functions we see
that the analytic form  of $\gc_{n\kk}^{\(e-p,e/h\)}$ are proportional to two terms appearing in Eq.\ref{eq:FAN_9}. 
Nevertheless, as we will clearly see in the case of bulk silicon, the dynamics induced by the KBE equation is a cascade process where many phonons are 
emitted or adsorbed and the final decay time of the electronic levels will be very different from the equilibrium case.

By using approximations (vi) and (vii) the e--e kernel, Eq.\ref{eq:KBE_13}, can be worked out in a way similar to the e--p case. Although the math
is more involved the final outcome is quite similar to the e--p case. Indeed it turns out that
\begin{align}
\left.
S_{n\kk}^{e-e,<}\(T;\gb\)\right|_{CCA}=-\xi_{n\kk}^{\(e-e,h\)}\(T;\gb\)\[2-f_{n\kk}\(T;\gb\)\]+\xi_{n\kk}^{\(e-e,e\)}\(T;\gb\)f_{n\kk}\(T;\gb\).
\label{eq:CCA_10}
\end{align}
The two additional lifetimes $\xi_{n\kk}^{\(e-e,e/h\)}$ are defined to be:
\begin{multline}
\xi_{\nk}^{e-e,e}\(T;\gb\)=4\(N\Omega\)^{-2}\sum_{\qq \pp}\sum_{m_1 m_2 m_3}
\left|W^{\qq}_{\substack{n m_1 \kk \\ m_3 m_2 \pp }}\right|^2 \\
R^{\qq}_{\substack{n m_1 \kk \\ m_3 m_2 \pp }} \(2-f_{m_1\kmq}\(T;\gb\)\)\(2-f_{m_3\pp}\(T;\gb\)\)f_{m_2\pmq}\(T;\gb\),
\label{eq:CCA_11}
\end{multline}
and
\begin{multline}
\xi_{\nk}^{e-e,h}\(T;\gb\)=4\(N\Omega\)^{-2}\sum_{\qq \pp}\sum_{m_1 m_2 m_3}
\left|W^{\qq}_{\substack{n m_1 \kk \\ m_3 m_2 \pp }}\right|^2 \\
R^{\qq}_{\substack{n m_1 \kk \\ m_3 m_2 \pp }} f_{m_1\kmq}\(T;\gb\)f_{m_3\pp}\(T;\gb\)\(2-f_{m_2\pmq}\(T;\gb\)\).
\tag{\ref{eq:CCA_11}$'$}
\end{multline}
We have introduce the function $R$ defined as:
\begin{multline}
R^{\qq}_{\substack{n m_1 \kk \\ m_3 m_2 \pp }}=
\[\lb\Im E_{m_1\kmq}\(\gb\)\rb+\lb\Im E_{m_3\pp}\(\gb\)\rb+\lb\Im E_{m_2\pmq}\(\gb\)\rb+\lb\Im E_{n\kk}\(\gb\)\rb\]\\
\{\[\Re\(E_{m_1\kmq}\(\gb\)+E_{m_3\pp}\(\gb\)-E_{m_2\pmq}\(\gb\)-E_{n\kk}\(\gb\)\)\]^2+\\
\[ \lb\Im E_{m_1\kmq}\(\gb\)\rb+\lb\Im E_{m_3\pp}\(\gb\)\rb+\lb\Im E_{m_2\pmq}\(\gb\)\rb+\lb\Im E_{n\kk}\(\gb\)\rb\]^2\}^{-1}.
\label{eq:CCA_12}
\end{multline}
The physical interpretation of the e--e assisted carrier dynamics is very similar to the e--p case. The only difference is that, instead of phonons, 
the carrier decay is due to the creation of e--h pairs. The basic transition that contributes to the e--e scattering is a four bodies
process: the initial electron (or hole) scatters in an e--h pair plus another electron (or hole). If we distinguish between the bath
electrons and the optically pumped ones (the carriers) we see that these four bodies can, quite generally, be distributed in both group of states.

Nevertheless in the case of semiconductors and insulators it is meaningful to distinguish between 
intra--carrier and extra--carrier channels. In the first case the e--h pair is created among the optically pumped carriers while in the second case the pair is 
created across the gap involving the bath electrons. Of course these two channels, in general, contribute at the same time to the overall dynamics. But, as we
will see in Sec.\ref{sec:silicon} their relative strength depends on the carrier energy.

In conclusion the total rate equation governing the carrier dynamics that will be used in the Sec.\ref{sec:silicon} is given by Eq.\ref{eq:KBE_14}
with $\bSS^{<}$ defined by Eqs.\ref{eq:CCA_4} and \ref{eq:CCA_10}. I coded this equation in the Yambo code\cite{AndreaMarini2009}, in the same
numerical framework where the equilibrium GW and Fan self--energies are coded. The existing structure of Yambo has already produced several
publications and it provides, therefore, the most appropriate environment to perform out--of--equilibrium simulations in a
controllable manner. The e--p matrix elements and phonon modes are calculated, instead, with the Quantum Espresso package\cite{Giannozzi2009}.

\subsection{A detailed balance condition to ensure that the number of electrons is conserved}
In the case of e--e scattering the actual possibility of creating high energy electron--hole\,(e--h) pairs can modify, during the real--time
dynamics, the number of carriers that fill the conduction bands with respect to the pairs initially photo--excited. 
This process is particularly evident for carriers created with energy larger than $2E_g$ below the VBM, or $2E_g$ above the CBM.

In the e--p case, instead, the maximum available energy that any carrier can dissipate during the relaxation is the Debye energy that, in the
systems I consider in this work, is much smaller than the gap. In this case, indeed, it is important that the theory preserves the
correct number of photo--excited electrons during the whole simulation.

I use this example to discuss a methodological aspect embodied in the use of a fully \ai framework. Indeed the e--p matrix elements
$g^{\qq \gl}_{n' n \kk}$, defined in Eq.\ref{eq:FAN_7}, can be written as
\begin{align}
g^{\gql}_{n' n \kk}\equiv g^{\kk-\pp \gl}_{n' n \kk}=\sum_{s \ga} \(2 M_s \go_{\kk-\pp\gl}\)^{-1/2} e^{i\(\kk-\pp\)\cdot\tau_s} 
\la n'\pp |\frac{\partial \Vscf^{\(s\)}\(\rr\)}{\partial{R_{s\ga}}} | n \kk \ra_{0} \xi_{\ga}\(\kk-\pp, \gl|s\).
\label{eq:DB_1}
\end{align}
We can, now, consider the KBE for the two isolate states $|n\kk\ra$ and $|m\pp\ra$ that are scattered one into the other 
by means of the $S_{n\kk}^{e-p,<}\(T;\gb\)$. For simplicity I only consider the dynamics induced by the e--p kernel at zero temperature and for a single
phonon band $\mu$. This dynamics
can be rewritten in terms of four lifetimes: $\gc_{n\kk}^{\(e/h\)}, \gc_{m\pp}^{\(e/h\)}$:
\begin{gather}
\gc_{n\kk}^{\(h\)}\(T\)=  \frac{\lb g^{\kk-\pp,\mu}_{nm\kk} \rb^2 }{N} P^{\(abs\)}_{nm\kk\(\kk-\pp\)\mu}f_{m\pp}\(T\),
\label{eq:DB_2}\\
\gc_{n\kk}^{\(e\)}\(T\)=  \frac{\lb g^{\kk-\pp,\mu}_{nm\kk} \rb^2 }{N} P^{\(emit\)}_{nm\kk\(\kk-\pp\)\mu}\(2-f_{m\pp}\(T\)\),
\tag{\ref{eq:DB_2}$''$}\\
\gc_{m\pp}^{\(h\)}\(T\)=  \frac{\lb g^{\pp-\kk,\mu}_{mn\kk} \rb^2 }{N} P^{\(abs\)}_{mn\pp\(\pp-\kk\)\mu}f_{n\kk}\(T\),
\tag{\ref{eq:DB_2}$''$}\\
\gc_{m\pp}^{\(e\)}\(T\)=  \frac{\lb g^{\pp-\kk,\mu}_{mn\kk} \rb^2 }{N} P^{\(emit\)}_{mn\pp\(\pp-\kk\)\mu}\(2-f_{n\kk}\(T\)\).
\tag{\ref{eq:DB_2}$'''$}
\end{gather}
Now we notice that, thanks to simple symmetry arguments it can be shown that 
\begin{align}
g^{\pp-\kk,\mu}_{mn\kk}= \[ g^{\kk-\pp,\mu}_{nm\kk}\]^*.
\label{eq:DB_3}
\end{align}
Moreover from Eqs.\ref{eq:CCA_5}--\ref{eq:CCA_6} it follows that $P^{\(emit\)}_{nm\kk\(\kk-\pp\)\mu}=P^{\(abs\)}_{mn\pp\(\pp-\kk\)\mu}$ 
and, consequently, from Eqs.\ref{eq:DB_2} we see that:
\begin{multline}
\frac{\partial}{\partial T} f_{n\kk}\(T\)= \frac{\lb g^{\kk-\pp,\mu}_{nm\kk} \rb^2 }{N} P^{\(emit\)}_{nm\kk\(\kk-\pp\)\mu}\(2-f_{m\pp}\(T\)\)
\(2-f_{n\kk}\(T\)\)\\-\frac{\lb g^{\kk-\pp,\mu}_{nm\kk} \rb^2 }{N} P^{\(abs\)}_{nm\kk\(\kk-\pp\)\mu}f_{m\pp}\(T\)f_{n\kk}\(T\),
\label{eq:DB_4}
\end{multline}
and
\begin{multline}
\frac{\partial}{\partial T} f_{m\pp}\(T\)= \frac{\lb g^{\kk-\pp,\mu}_{nm\kk} \rb^2 }{N} P^{\(abs\)}_{nm\kk\(\kk-\pp\)\mu}\(2-f_{m\pp}\(T\)\)
\(2-f_{n\kk}\(T\)\)\\-\frac{\lb g^{\kk-\pp,\mu}_{nm\kk} \rb^2 }{N} P^{\(emit\)}_{nm\kk\(\kk-\pp\)\mu}f_{m\pp}\(T\)f_{n\kk}\(T\).
\label{eq:DB_5}
\end{multline}
By summing Eq.\ref{eq:DB_4} and Eq.\ref{eq:DB_5} one gets 
\begin{multline}
\frac{\partial}{\partial T} \(2-f_{m\pp}\(T\)-f_{n\kk}\(T\)\)= \\
-\frac{\lb g^{\kk-\pp,\mu}_{nm\kk} \rb^2 }{N} \[P^{\(abs\)}_{nm\kk\(\kk-\pp\)\mu}+P^{\(emit\)}_{nm\kk\(\kk-\pp\)\mu}\]
\(2-f_{m\pp}\(T\)-f_{n\kk}\(T\)\).
\label{eq:DB_6}
\end{multline}
But from the boundary condition $\(2-f_{m\pp}\(T=0\)-f_{n\kk}\(T=0\)\)=0$ it follows $\frac{\partial}{\partial T}
\(2-f_{m\pp}\(T\)-f_{n\kk}\(T\)\)=0$. The total number of electrons in the two levels made scattering by the e--p interaction is conserved, as it
should be.

Eq.\ref{eq:DB_3} thus represents a sort of detailed balance condition that ensures that the fraction of charge lost by a given
state $|n\kk\ra$ is exactly compensated by the charge added to the $|m\pp\ra$  state. In total there is no nonphysical charge lost or increase.
Of course the level of accuracy of this detailed balance condition is ensured by the numerical precision under which Eq.\ref{eq:DB_3} is
satisfied. This precision must be carefully checked and, eventually imposed, as on long--time simulations (of the order of picoseconds) tiny
deviations can induce important deviations of the total number of electrons conservation.

\section{An application in a paradigmatic solid: bulk silicon}
\label{sec:silicon}
After all the math and approximations introduced in the previous sections it is instructive to apply all the machinery composed by
the KBE's written in the \ai framework to perform realistic simulations. The paradigmatic case I consider is bulk silicon. This a 
quite simple material that, nevertheless, has an interesting property: the minimal gap is indirect.

\begin{figure}[h]
\begin{center}
\epsfig{figure=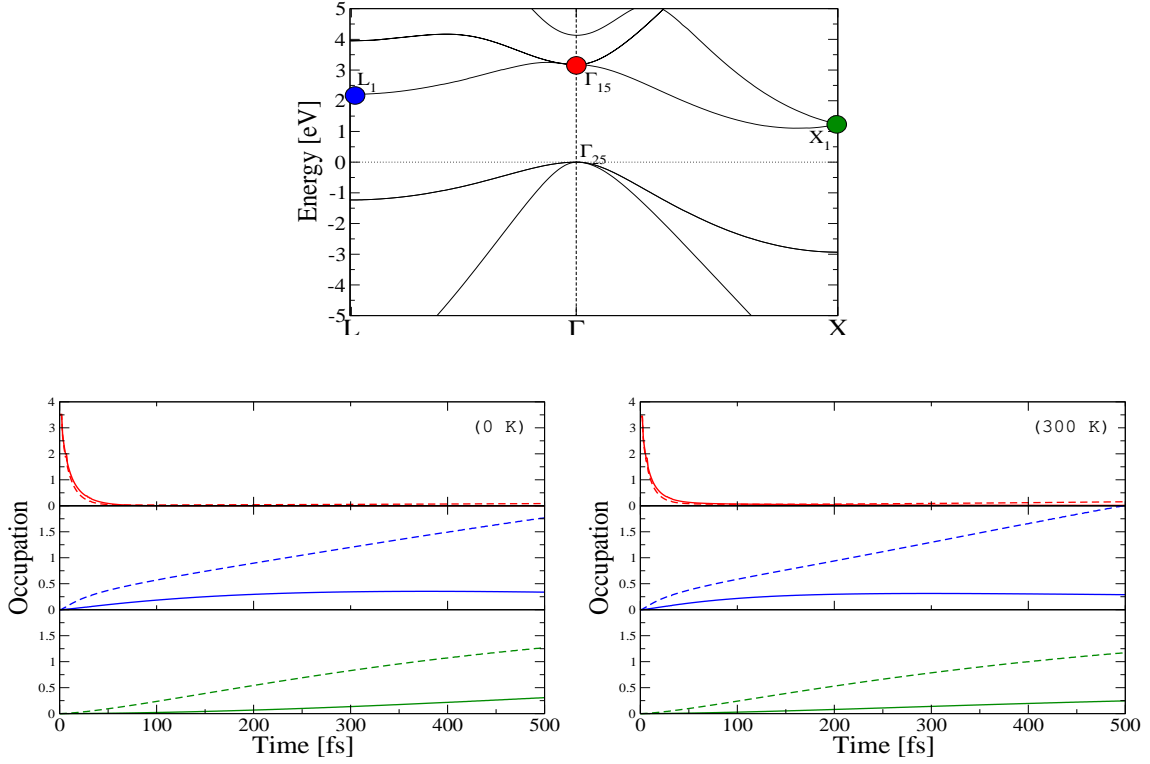,clip=,bbllx=49,bblly=509,bburx=513,bbury=825,width=15cm}
\end{center}
\caption{\footnotesize{
Time dependent occupations (lower frames) $f_{\nk}\(T;\gb\)$ as a function of the $T$ for two external fixed temperature ($0\,K$ left frame and
$300\,K$ right frame) for 
three selected states in the band--structure of Silicon (upper frame). At $T=0$ only the $\Gamma_{15}$ is artificially populated with 4 electrons 
needed to fill 2 of the 3 energy degenerated states. This explains why the y--axis goes from 0 to 4.
In the lower frames the dynamics is done including only e--p scattering (continuous lines) and e--e plus e--p scattering (dashed lines).
In this case, as shown in Fig.\ref{fig:EQ_linewidths}, the energy of the $|\Gamma_{15}\ra$ state is below the $2E_g$ threshold. This means, indeed, that
at $T=0$ the e--p scattering is largely dominant over the e--p one. But, as the time increases and the initial charge is fragmented in lower 
energy states the intra--carriers e--e scattering quickly increases leading to a much faster population of the $X_1$ state.
}}
\label{fig:GAMMA_occs}
\end{figure}

From the point of view of the carrier dynamics this is an important aspect. Indeed the optical gap is located at the $\Gamma$ point and it is 
larger ($\sim\,3.2\,eV$) of the indirect gap ($\sim\,1.2\,eV$). Thus if an electron is optically pumped in the conduction states this will populate
the energy region surrounding the CBM, that is the $\Gamma_{15}$ state, and the relaxation is
expected to occur along the $\Gamma X$ line (see the upper frame of Fig.\ref{fig:GAMMA_occs}) towards the $X_1$ point.

This has been, indeed, experimentally measured\cite{Ichibayashi2011} by using Time--Resolved Two--Photon Photoemission\,(2PPE) 
and the time--decay of a level measured along the $\Gamma X$ has been found to be around 40\,fs.

I do not want, however, to interpret and reproduce the 2PPE experiment but to use bulk silicon to illustrate the relative balance of the 
e--e and e--p channels in the carrier dynamics on the basis of the initial carrier energy and density. I will also briefly discuss the effect of
the scattering with acoustic phonons keeping in mind the large number of simulations done using simple optical--only phonon models.

The simulation is performed by defining an {\it ad--hoc} initial condition for the occupations and let the system evolve by means of 
Eq.\ref{eq:KBE_1}. In the first series of calculations I take
\begin{align}
f_{\nk}\(t=0;\gb\)=2\quad\text{when}\quad |\nk\ra=|\Gamma_{15}\ra.
\label{eq:silicon_1}
\end{align}
Two electrons (to account for the two spin channels) are moved by hand in two out of the three states that are energetically degenerate with the $\Gamma_{15}$ stat. This corresponds to an
initial carried density of $3.2\times\,10^{21}\,cm^{-3}$. 
As the  $\Gamma_{15}$ state is at the border of the $2 E_g$ threshold its extra--carriers contribution to the e--e lifetime is very small. This is because the energy 
of the state is not enough to create e--h pairs across the minimum gap. The real--time occupations of three paradigmatic states: $|\Gamma_{15}\ra$,
$|L_1\ra$ and $|X_1\ra$ are showed in Fig.\ref{fig:GAMMA_occs}. In the $\kk$--grid I have used the $X_1$ state is the nearest to the CBM.

In Fig.\ref{fig:GAMMA_occs} the functions $f_{\Gamma_{15}}\(T;\gb\)$ (red curves),  $f_{L_{1}}\(T;\gb\)$ (blue curves) and  
$f_{X_1}\(T;\gb\)$ (green curves) are showed comparing the dynamics with only the e--p scattering (continuous lines) and
with the e--p and the e--e scatterings (dashed lines) included. We immediately see some clear trends. The decay if the $\Gamma_{15}$ is very fast, with
an exponential time--dependence with
a lifetime of the order of $12\,fs$. This value is very similar to the value of $16\,eV$ obtained by calculating
$\Gamma^{e-p,eq}_{\Gamma_15}\(\gb\rightarrow\infty\)$ showing that the e--e contribution is practically zero.  As the time increases the initial charge
is fragmented and lower state are populated. At this point the intra--carriers contribution to the e--e is rapidly increased  and the contribution to the filling
is largely underestimated.

The message is thus clear. In the very short time the dynamics of carriers with energy below the $2E_g$ threshold is e--p dominated. Nevertheless in
this regime the use of the CCA is highly questionable and a more careful full--memory treatment is required. As time increases also in this 
energy range the intra--carriers contribution to the carrier dynamics makes the e--e scattering dominant compared to the e--p channel.
\vspace{.2cm}
\begin{figure}[h]
\begin{center}
\epsfig{figure=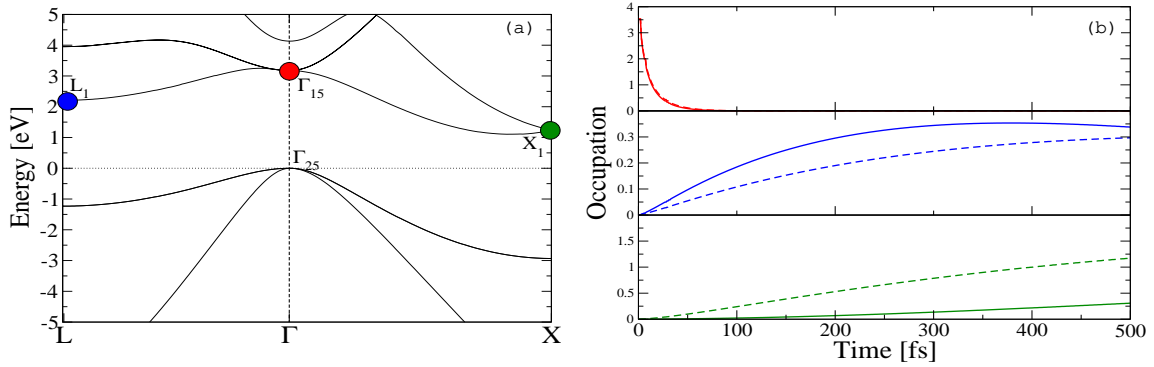,bbllx=27,bblly=658,bburx=490,bbury=808,width=15cm}
\end{center}
\caption{\footnotesize{
Time dependent occupations (right frame) $f_{\nk}\(T;\gb\)$ as a function of the $T$ for an external vanishing temperature for 
three selected states in the band--structure of Silicon (left frame). At $T=0$ only the $\Gamma_{15}$ is artificially populated with 4 electrons 
needed to fill 2 of the 3 energy degenerated states.
In the right frame the dynamics is done only including the e--p scattering but including only optical phonon modes (continuous lines) or all modes (dashed lines).
We see that the effect of the acoustic modes is not at all negligible, especially for the $X_1$ state filling.
}}
\label{fig:GAMMA_ONLY_OPT}
\end{figure}

An important aspect that must be taken in account in the analysis of the time--dependencies showed in Fig.\ref{fig:GAMMA_occs} is that the overall 
filling of the $X_1$ and $L_1$ states is enhanced when the e--e scattering is included also because of the promotion of additional electrons
from the valence to the conduction by the creation of e--p pairs across the gap. This process, although energetically impossible for the $\Gamma_{15}$ state, is made
possible by the energy indetermination caused by the use of the QP approximation for the retarded Green's function. Indeed thanks to the Lorentzian
line--shape, the $R$ functions (see Eq.\ref{eq:CCA_12}) can be non--zero for arbitrary values of the energy levels involved in the e--e scattering. This is, in my opinion, a
stringent limitation imposed by the QP approximation that can be easily avoided by using more refined approximations or by calculating the equilibrium
spectral functions numerically in the GW and Fan approximations.

By looking at the two lower frames of Fig.\ref{fig:GAMMA_occs} we see that the increase of the temperature from $0$ to $300$ K does not
appreciably change the occupation dynamics in the first 500\,fs. This is reasonable as the e--e lifetimes are not explicitly temperature
sensitive and, as far as the e--p channel is considered, the Silicon Debye temperature is as large as 
$645$ K. Thus at room temperature only the low--energy acoustic modes are populated. Although those contribute for half of the e--p lifetime 
the overall effect on the dynamics is marginal. 

\begin{figure}[h]
\begin{center}
\epsfig{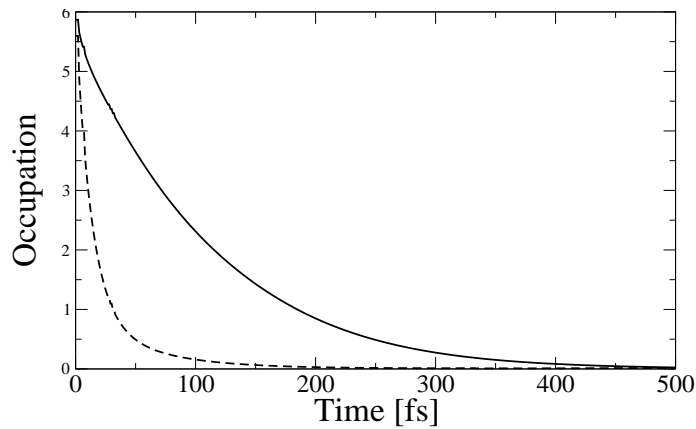}
\end{center}
\caption{\footnotesize{
Time dependent occupation  as a function of the $T$ for the set of electronic states with energy $5.76$\,eV, well above the 
$2E_g$ threshold energy. The simulation is performed with the e--p scattering and including (dashed line) and not including (continuous line)
the e--e contribution. For this high energy level the equilibrium e--e and e--p lifetimes are similar  and both contribute to the 
decay of the state. This contrasts the $\Gamma_{15}$ case where the e--p scattering dominates in the short--time dynamics.
}}
\label{fig:k6b7}
\end{figure}

An important issue that can be fully exploited by the present calculations is, indeed, the role played by the acoustic phonon modes that, in 
model calculations, are often neglected. As shown in Fig.\ref{fig:GAMMA_ONLY_OPT} the acoustic modes, even at zero temperature, contributes for more
than half of the filling of the $X_1$ state. Their contribution to the $\Gamma_{15}$ state is, instead, minor and it slowly increase as the energy of the state
approaches the CBM. This phenomena is connected with the physics of both the e--e and e--p induced relaxation of the carriers. This is indeed
a cascade process where the initial $\Gamma_{15}$ electrons jump towards other states by the emission of, initially, a single phonon and a single e--h pair. This
elementary process is very fast, of the order of even faster of $10$\,fs. But in order to reach the $X_1$ state the electron undergoes several 
of these elementary processes and the overall relaxation is a much slower process.

The last aspect I want to analyze by using bulk Silicon as paradigmatic system is the relative influence of the e--e and e--p scattering channels in
the short--time regime for carriers pumped well above the $2E_g$ threshold. To this end I change the boundary condition give by Eq.\ref{eq:silicon_1}
to be
\begin{align}
f_{\nk}\(t=0;\gb\)=2\quad\text{when}\quad E_{\nk}=5.76\,eV
\label{eq:silicon_2}
\end{align}
This corresponds to an initial carried density of $1.9\times\,10^{22}\,cm^{-3}$, an order of magnitude larger then in the
$\Gamma_{15}$ case. From Fig.\ref{fig:EQ_linewidths} we see that the equilibrium QP e--e and e--p lifetimes are very similar. And,
indeed, the decay of the $5.76\,eV$ state is very much affected by the e--e scattering that, in this case, at short times is dominated
by the extra--carriers channel.

\section{Conclusions}
\label{sec:conclusions}
In conclusion I have presented an \ai approach to the solution of the Baym--Kadanoff equations in realistic materials.
The present approach is based on two major approximations: the generalized Baym--Kadanoff ansatz and the 
completed collision approximation. These makes the calculation feasible, as shown in the case of bulk silicon
and provide a consistent and solid theoretical and numerical framework that can be applied to
any kind of material that can be studied using the standard \ai Many--Body Perturbation Theory
techniques.
Indeed, a major point of the present approach is that all the presented theoretical and numerical tools has been coded
in an existing long--term, stable and highly productive code: the Yambo code\cite{AndreaMarini2009}.
By using bulk silicon as a paradigmatic material I show that, in semiconductors and insulators, the presence of an electronic
gap makes the e--p channel dominant over the e--e one in the short--time regime ($T\lesssim 100$\,fs) and
for low energy carriers injected or optically pumped above the conduction band minimum but below the $2E_g$ threshold. This is due to the fact that one of the 
most important kinematic channels in the e--e scattering is the one that involves e--h pairs created across the energy gap, and those pairs
are energetically forbidden when the carriers are pumped with energy that is  $2E_g$ below the CBM.
It is important to stress that on the basis of the relative balance of the e--e and e--p scattering channels presented in
this work it is not possible be draw general conclusions for all kind of materials. 
If on one side I am convinced that, even if the CCA weakens the results 
obtained in the very--short time scale, the arguments based on the energy of the initial carrier to motivate a potential 
bigger importance of the e--p channel is reliable for any kind of semiconductor or insulator. Of course the same argument
does not apply to metals. On the other hand silicon is not a system known for its strong e--p coupling. There are systems like
bulk Diamond\cite{Giustino2010}, carbon nano--structures\cite{cannuccia} or super--conductors\cite{supercond} where the e--p scattering is enormously strong and 
this can strengthen the e--p channel over the e--e one.

\section*{References}

\end{document}